\documentclass[]{jkas} 
\usepackage[table]{xcolor}
\usepackage{algorithm}
\usepackage{algpseudocode}
\usepackage{caption}

\def\year{2023} 
\def\volume{56} 
\def\issue{1} 
\def\beginpage{1} 
\def\received{---} 
\def\accepted{---} 
\def\published{---} 
\date{Received \received; Accepted \accepted; Published \published}

\title{%
Masking Algorithm for CCD Bleeding in Korea Microlensing Telescope Network Images
}

\author[1]{Jiseop Shin}{0009-0005-3944-1457}
\author[1]{Mankeun Jeong}{0009-0003-1280-0099}
\author[1,$\star$]{Myungshin Im}{0000-0002-8537-6714}
\author[1]{Seo-Won Chang}{0000-0002-3118-8275}
\author[2]{Sang-Mok Cha}{0000-0002-7511-2950}
\author[1]{Seong-Kook Lee}{0000-0001-5342-8906}

\affil[1]{SNU Astronomy Research Center, Astronomy Program, Department of Physics and Astronomy,
Seoul National University, 1 Gwanak-ro, Gwanak-gu, Seoul 08826, Republic of Korea}
\affil[2]{Korea Astronomy and Space Science Institute, Daejeon 34055, Republic of Korea}

\def\corrauthor{%
M. Im, \email{myungshin.im@gmail.com}
}

\def\runningauthor{%
Shin et al.
}

\def\runningtitle{%
Masking Algorithm for CCD Bleeding
}

\def\keywords{%
telescopes --- surveys --- instrumentation: detectors --- techniques: image processing
}

\def\abstracttext{%
Astronomical CCD images are often affected by bleeding from bright sources, producing column-aligned streaks that degrade source detection and photometry. We present a pixel-level masking algorithm for bleeding in Korea Microlensing Telescope Network (KMTNet) images, in which these streaks extend along detector columns away from the readout register. The algorithm identifies bleed-generating pixels using a conservative bleeding threshold and a bleeding index that quantifies excess signal along the expected bleeding direction. The extent of each bleed trail is then determined by termination criteria based on a background-relative detection threshold and a continuity condition. To optimize these termination parameters, we generate multiple candidate masks and evaluate them with an F$_\beta$-like score that favors high purity while preserving recovery of sources whose photometry is biased by bleeding. Using the final mask, we apply interpolation-based cleaning and assess the performance in terms of completeness and photometric accuracy in a crowded KMTNet field, using Gaia DR3 as a reference. The source completeness improves from 94.08\% to 98.61\%, primarily by enabling proper aperture placement that was previously hindered by bleeding. For selected bleeding-affected sources, the root-mean-square photometric offset decreases from 0.623 mag to 0.068 mag. These results demonstrate that the algorithm provides a practical framework for mitigating bleeding artifacts in KMTNet images. We further examine the stability of the algorithm across different observing conditions and discuss limitations of the current masking and cleaning procedure.
}

\begin{document}
\jkashead 


\section{Introduction}\label{sec:intro}

The Charge-Coupled Device (CCD) has played a central role in the development of modern observational astronomy. It converts incident photons into electrical charge and stores it in individual pixels. The accumulated charge is then transferred sequentially across the detector toward the readout register, typically first along columns and then along rows. Owing to its high quantum efficiency, broad dynamic range, low read noise, excellent linearity, and photometric stability, the CCD has become a standard detector for astronomical imaging and precision photometry. However, despite these advantages, it is also subject to several characteristic artifacts, especially those aligned with the charge-transfer direction.

The most familiar column artifact is blooming (also called bleeding or a bleed trail in some studies; \citealt{stankiewicz2008, coupon2018, desai2016, morganson2018}), in which excess charge from a saturated pixel overflows into neighboring pixels along the column. Because this overflow occurs during exposure when the accumulated charge exceeds the pixel full-well capacity, excess charge spills along the detector column. Blooming has been documented in a wide variety of instruments and surveys, including the Hubble Space Telescope, Hyper Suprime-Cam on the Subaru Telescope, and the Dark Energy Camera used for the Dark Energy Survey \citep{stankiewicz2008, marinelli2025, coupon2018, morganson2018, desai2016, winecki2024}. A second class is the charge-transfer-efficiency (CTE) trail. Cumulative radiation damage, such as that caused by cosmic rays, creates localized charge traps in the silicon lattice. During readout, these traps capture a fraction of the transferred signal charge and release it with a finite delay during subsequent clocking shifts, depositing charge into trailing pixels and producing a one-sided streak that extends away from the readout register. Unlike blooming, CTE trails can be produced by any pixel with degraded CTE, regardless of count level. A third class of artifact is associated with residual charge from bright or saturated sources that remains within the CCD after the initial exposure or readout. The physical location of this residual charge can depend on the detector architecture and clocking configuration. In the Pan-STARRS1 (PS1) GPC1, \citet{waters2020} described burn trails as arising when charge from nearly saturated pixels remains in the undepleted region of the silicon and gradually leaks out during readout. In the LSST Camera e2v CCDs, \citet{polin2025} showed that, under certain parallel clocking voltage configurations, charge can reach the Si--SiO$_2$ interface and become trapped in long-lived surface states. In both cases, delayed release of residual charge from saturated sources can produce extended one-sided trails or residual images, with the severity depending strongly on the clocking configuration.

\begin{figure}[t]
\centering
\includegraphics[width=\linewidth]{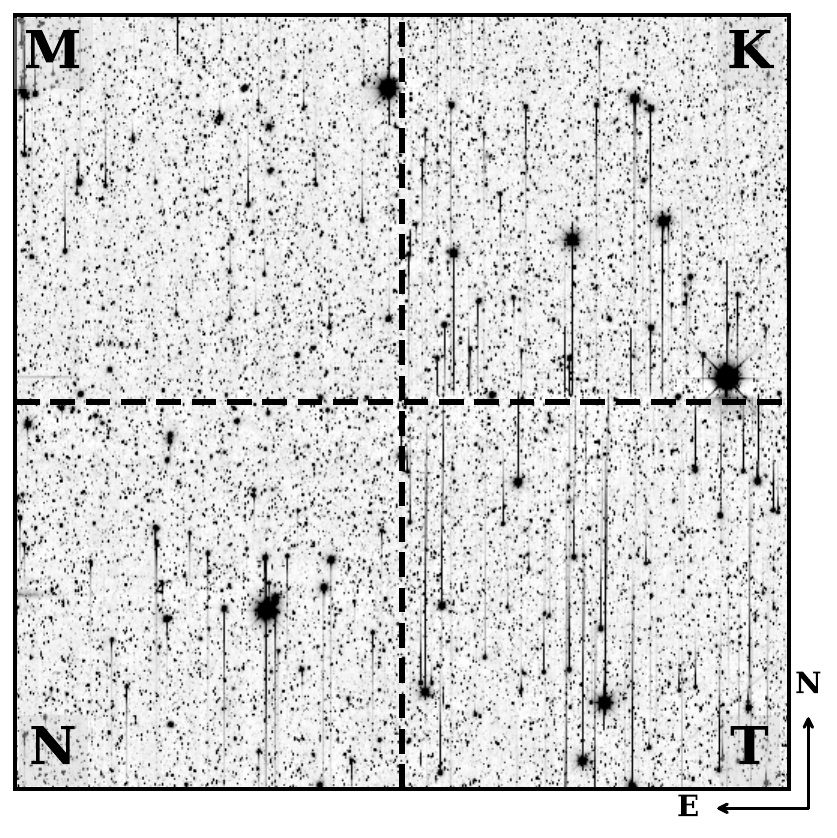}
\caption{Stacked $I$-band image of field 1084 in KS4, obtained at CTIO. Dashed lines indicate the boundaries of the four CCD chips. The bleed trails extend upward in the M and T-chips and downward in the K and N-chips, away from the readout registers.} \label{fig:bleeding image}
\end{figure}

This type of artifact is also visible in the Korea Microlensing Telescope Network (KMTNet; \citealt{kim2016}) (Figure~\ref{fig:bleeding image}). KMTNet is a global network of three identical 1.6-m telescopes located at the Cerro Tololo Inter-American Observatory in Chile (CTIO), the South African Astronomical Observatory in South Africa (SAAO), and the Siding Spring Observatory in Australia (SSO). Each telescope employs a mosaic detector composed of four $9\,\mathrm{k}\times9\,\mathrm{k}$ wide-field CCD chips (e2v CCD290-99), providing a total field of view of $2\times2~\mathrm{deg}^{2}$. The four chips are labeled K, M, T, and N, and each chip has eight readout channels. The readout registers are located at the top of the K- and N-chips and at the bottom of the M- and T-chips, causing the trails to appear in different directions depending on the chip location (Figure~\ref{fig:bleeding image}). Suppression of these charge-transfer artifacts in the e2v CCD290-99 depends sensitively on the clocking configuration. Tests by the KMTNet operations team indicate that these trails can be reduced by modifying the substrate voltage or the parallel-clock voltage levels (Cha et al., in preparation); however, such adjustments are not readily feasible with the current KMTNet controller configuration. These defects were already present at the earliest stage of KMTNet operations (see Figure 5 of \citealt{kim2016}). Although physically distinct from classical blooming, these features have been referred to as bleeding, bleed trails, or blooming in the KMTNet literature \citep{kim2016, zhang2023, yang2024mnras, paek2025, viana2025}. In this paper, we follow this convention and refer to them collectively as bleeding or bleed trails.

While wide longitudinal coverage of KMTNet has enabled a broad range of time-domain studies, including microlensing \citep{zang2021, zhang2023, shin2016, hwang2022, gould2022}, supernovae \citep{afsariardchi2019, ni2022}, variable stars \citep{chang2018, kim2025}, comets \citep{paek2026}, asteroids \citep{bach2019}, active galactic nuclei \citep{kim2018, kim2024}, gamma-ray bursts \citep{yang2024}, and gravitational-wave counterparts \citep{im2017, troja2017, kim2021, paek2024, paek2025}, no systematic method has been available to identify and mask bleeding-affected pixels at the image level. This posed two major problems. First, bleeding compromises photometric accuracy by artificially increasing the measured flux of objects located along the streak. For example, \citet{zhang2023} showed that a bleed trail can artificially brighten a microlensing light curve, making the affected data inconsistent with other data sets. Bleeding can also bias background estimation, thereby degrading both the photometry and its associated uncertainties. Second, bleeding hampers source detection by interfering with deblending and the proper placement of photometric apertures, reducing completeness. In difference imaging analysis, bleed trails interfere with source detection and increase the number of false-positive detections \citep{paek2025, viana2025}.

Previous studies have handled these artifacts by excluding saturated sources from analysis \citep{zhang2023, yang2024mnras}, removing false positives through visual inspection, or applying machine-learning methods within transient alert systems \citep{paek2025, lee2025, viana2025}. Although effective in specific contexts, these approaches do not offer a general and uniform framework for identifying bleed-contaminated pixels and flagging affected sources in large survey pipelines. For similar artifacts in other instruments, \citet{waters2020} modeled bleed trails in PS1 GPC1 with an empirical power-law fit, while \citet{polin2025} mitigated the effect by modifying the CCD clocking scheme of the LSST Camera. However, these methods are not directly applicable to KMTNet images, since the architecture of the KMTNet CCDs differs from that of the PS1 GPC1 and LSST cameras. These limitations emerged clearly during the preparation of the KMTNet Synoptic Survey of Southern Sky (KS4; Im et al., in preparation) and motivated the development of the algorithm presented in this work.

In this paper, we present a Python-based, pixel-level algorithm that identifies bleeding-affected pixels and generates a binary mask for KMTNet images. Using this mask, we correct the affected pixels through interpolation and demonstrate improvements in both source detection completeness and photometric accuracy. The algorithm and the resulting bleeding flags have been incorporated into the KS4 pipeline \citep{jeong2026} and are included in the KS4 Data Release 1 products (DR1; \citealt{chang2026}).

This paper is organized as follows. Section~\ref{sec:Data} describes the data used to develop and validate the algorithm. Section~\ref{sec:Methods} details the masking procedure and the selection of the optimal masking parameters. Section~\ref{sec:Results} demonstrates the improvement achieved by applying the bleeding mask. Section~\ref{sec:Discussion} discusses the stability of the algorithm and the limitations of both the algorithm and the current bleeding-cleaning procedure. Finally, Section~\ref{sec:Conclusion} summarizes our main findings.

\section{Data}\label{sec:Data}
The bleeding masks are constructed from individual preprocessed KMTNet images reduced with the Korea Astronomy and Space Science Institute (KASI) pipeline (\citealt{kim2009, kim2013}). The preprocessing steps include overscan subtraction, dark subtraction, flat-field correction, and cross-talk removal. We refer to these preprocessed frames as single-epoch images throughout this paper.

To develop and tune the masking algorithm, we selected the KS4 field 1084, centered at ${\rm RA}=179^\circ$ and ${\rm Dec}=-34^\circ$. This field is among the most crowded fields in the KS4 DR1 footprint, with approximately 190,000 sources detected by \texttt{Source Extractor} (hereafter \texttt{SExtractor}; \citealt{bertin1996}) with \texttt{DETECT\_THRESH}=1.0, and is consequently among the fields most severely affected by bleeding. Bleeding is most prominent in the $I$-band, where longer-wavelength photons penetrate deeper into the silicon, so we use this band to develop and optimize the algorithm parameters. The algorithm is then applied to all KMTNet bands ($BVRI$). For development, we use images obtained at CTIO on April 7, 2021; the stability of the algorithm across different sites, bands, and observing conditions is discussed in Section~\ref{sec:Discussion}.

The masking algorithm operates on individual single-epoch chip images. In the KS4 pipeline, however, source detection and photometry are performed on stacked mosaic images that combine the four chips into a single $2\times2~\mathrm{deg}^{2}$ field. The bleeding masks must therefore be mapped onto the same mosaic grid. Accordingly, we reprojected the individual chip-level masks and merged them into a master bleeding mask using \texttt{SWarp} \citep{bertin2010}, following the general stacking strategy described in \citet{jeong2026}. Figure~\ref{fig:bleeding image} shows the resulting stacked $I$-band mosaic of field 1084.

\section{Bleeding Masking Algorithm}\label{sec:Methods}

\begin{figure}[t]
    \centering
    \includegraphics[width=\linewidth]{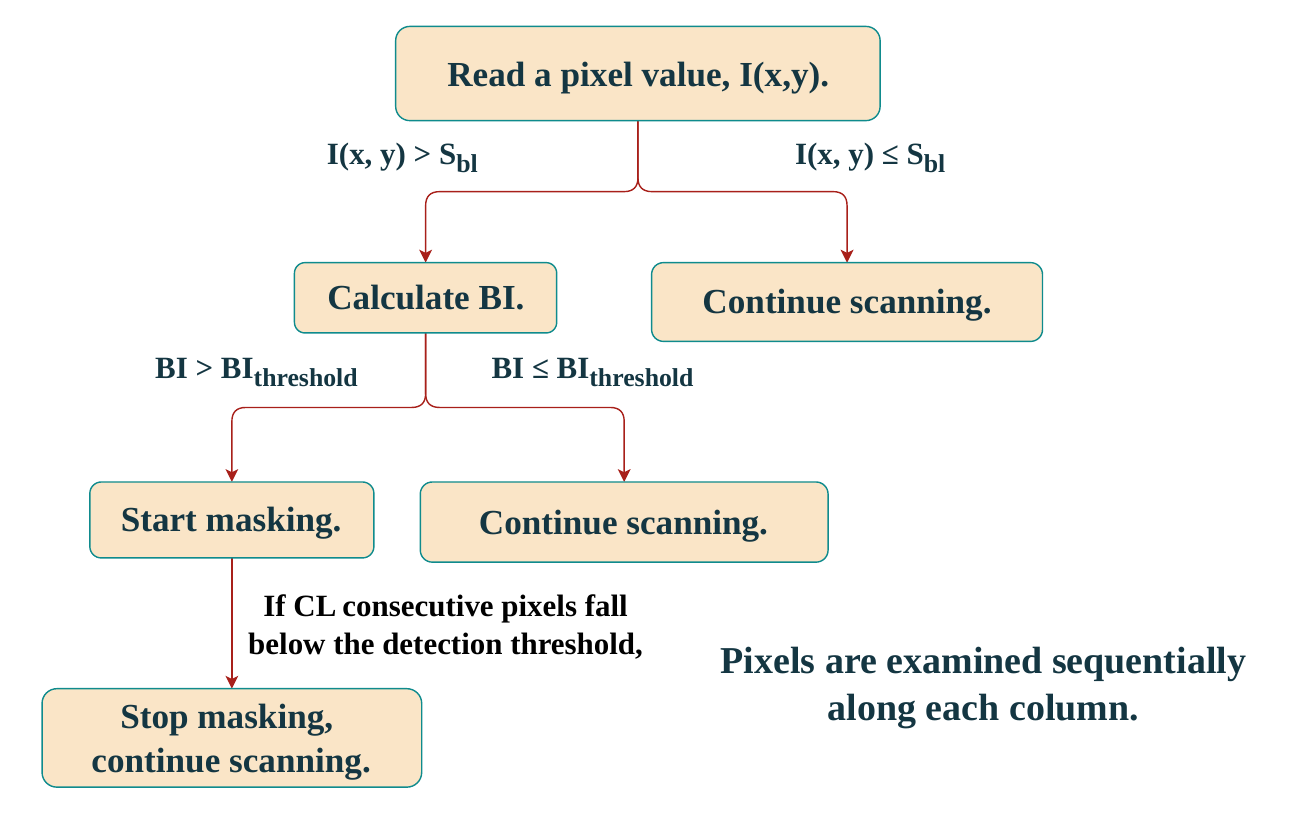}
    \caption{Flowchart illustrating the bleeding masking algorithm. Each parameter is explained in Section~\ref{sec:parameter selection}.}
    \label{fig:flowchart}
\end{figure}

The masking algorithm assigns a binary flag to each pixel: 1 for bleeding-affected and 0 for unaffected. As described in Section~\ref{sec:intro}, bleeding in KMTNet images originates from bright pixels and extends along the detector column away from the readout register. The algorithm uses this spatial pattern by scanning each column independently and performing two tasks: identifying the onset of each bleeding streak and determining where the streak terminates. Two parameters determine where the bleeding begins: the bleeding threshold, $S_{\rm bl}$, selects pixels likely to generate bleeding, and the bleeding index, $\mathrm{BI}$, distinguishes true bleed-generating pixels from other candidates. The termination of the mask is controlled by two additional parameters: the detection threshold, $n$, determines whether the pixel is affected by bleeding or not, and the continuity length, $CL$, sets the minimum required number of bleeding-unaffected pixels. The complete decision flow is summarized in Figure~\ref{fig:flowchart}.

Below, we define each parameter and describe how the adopted thresholds were chosen.

\subsection{Parameter Selection}\label{sec:parameter selection}

\subsubsection{Bleeding Threshold}\label{sec:bleeding threshold}
Candidate bleed-generating pixels are identified among very bright pixels; however, the precise pixel value at which bleeding begins is unknown. We therefore introduce a conservative threshold to identify candidate bleed-generating pixels, hereafter referred to as the bleeding threshold ($S_{\rm bl}$).

\begin{figure}[t]
    \centering
    \includegraphics[width=\linewidth]{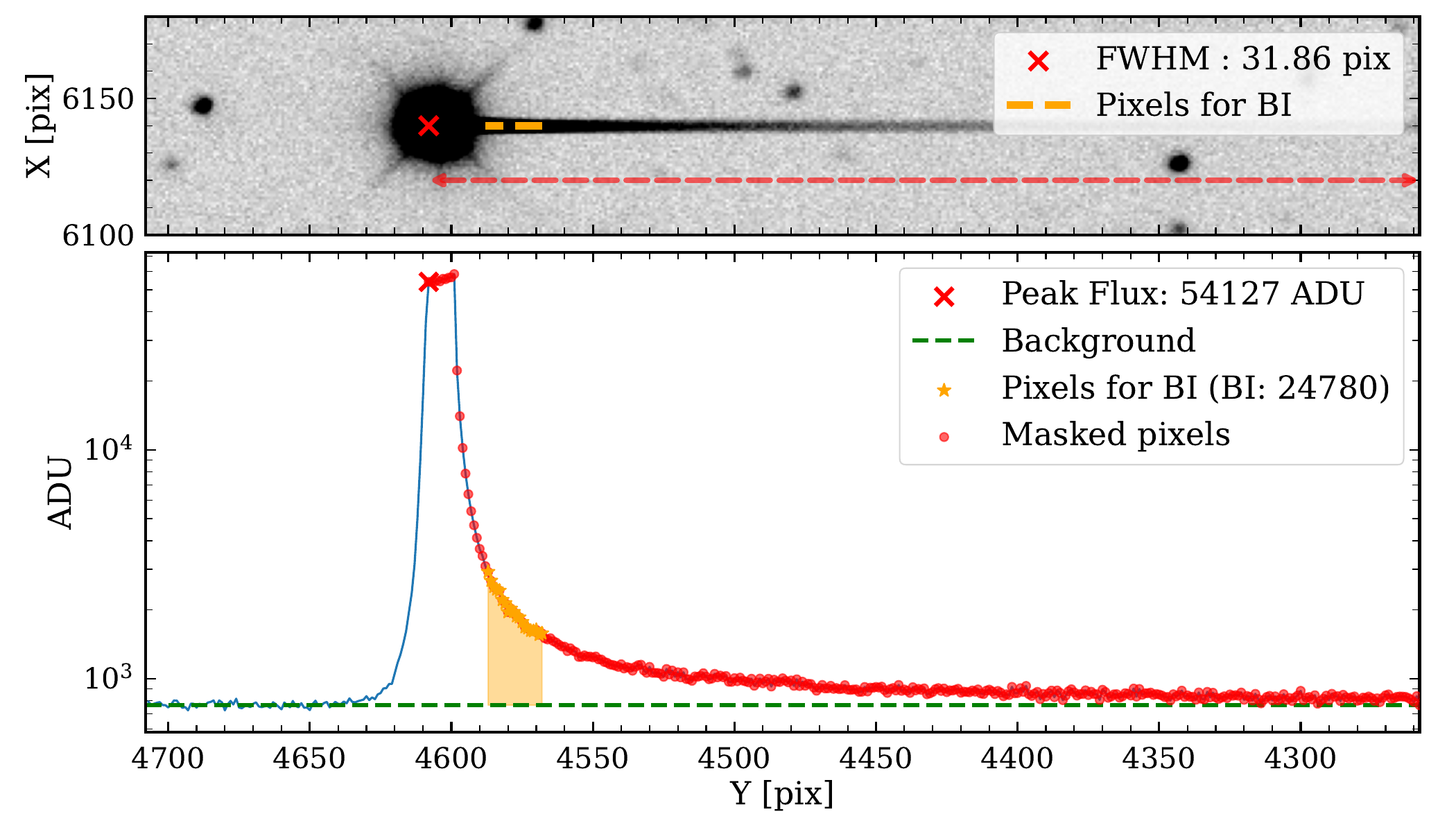}
    \includegraphics[width=\linewidth]{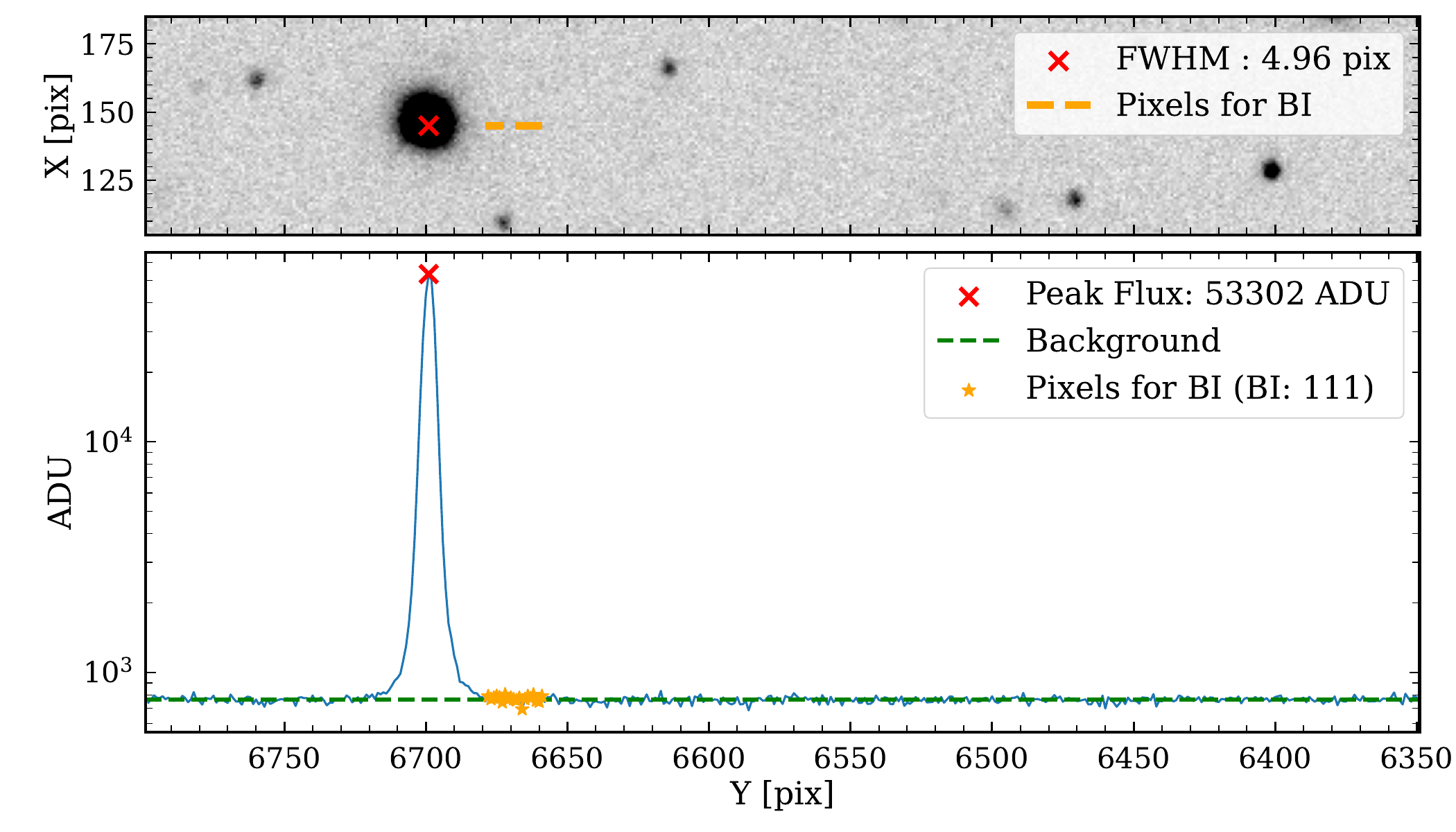}
    \caption{Images of two sources with bleeding (above) and without bleeding (below) are shown together with their ADU profiles along the column direction ($y$-axis). For each source, the bleed-generating pixel (peak flux) is marked with a red cross. The pixels used to compute $\mathrm{BI}$ are indicated by the yellow dashed line in the image and by yellow stars and a shaded region in the ADU profile. The green dashed line in the profile marks the background level. In the upper panel, pixels flagged by the bleeding mask (detection threshold $0.4\sigma$, continuity length 6; see Section~\ref{sec:termination criteria}) are highlighted with red circles in the profile and a red dashed line in the image.}
    \label{fig:ADU profile}
\end{figure}

To determine the threshold, we inspected images from all chips at all three sites and tested a range of trial values. We found that a threshold of 50,000 analog-to-digital units (ADU) successfully captures all pixels that generate bleeding. The only exception was the T chip at CTIO, where bleed trails begin to appear at around 45,000 ADU; for this chip, we therefore lowered the threshold to 40,000 ADU. We emphasize that this ADU cut is used only as a conservative criterion for identifying potential bleed-generating pixels, and that not every pixel above this threshold produces bleeding. Figure~\ref{fig:ADU profile} shows two bright sources whose peak values exceed 50,000 ADU; however, a bleed trail is present only in the upper panel.

\subsubsection{Bleeding Index}\label{sec:BI}

Because not all bright pixels identified in Section~\ref{sec:bleeding threshold} generate bleed trails, we introduce the bleeding index ($\mathrm{BI}$) as a quantitative diagnostic to distinguish genuine bleed-generating pixels from other bright pixels. The index measures the excess signal, relative to the local background, in the direction where bleeding is expected to propagate. For each candidate bleed-generating pixel, we consider a 20-pixel segment along the $y$-axis, beginning 20 pixels downstream of the pixel for the K- and N-chips, and 20 pixels upstream for the M- and T-chips. The $\mathrm{BI}$ is then defined as the background-subtracted sum of pixel values in this segment (Equation~\ref{eq:BI equation}). 
\begin{equation}\label{eq:BI equation}
    \mathrm{BI} = \sum_{i=1}^{20} \left(p_i - \mathrm{background}\right)
\end{equation}
Figure~\ref{fig:ADU profile} marks the candidate bleed-generating pixel with a red cross and highlights the 20-pixel segment used to compute $\mathrm{BI}$ with the yellow dashed line, yellow stars, and shaded region.

The choice of 20 pixels away from the selected pixel corresponds to $\sim4\times$ the median full width at half maximum (FWHM) of unsaturated point sources ($\sim 5$ pixels) that are unaffected by bleeding. We estimated this median FWHM using sources detected by \texttt{SExtractor}, applying a bleeding mask with provisional termination criteria (see Section~\ref{sec:termination criteria}). We then selected unsaturated sources with $\texttt{CLASS\_STAR}>0.9$, $\texttt{NIMAFLAGS\_ISO}<190$, and $\texttt{FLAGS}=0$. Here, \texttt{NIMAFLAGS\_ISO} is the \texttt{SExtractor} output parameter that counts the number of masked pixels within the isophotal aperture. We confirmed that the $\texttt{NIMAFLAGS\_ISO}<190$ criterion effectively rejects sources contaminated by bleeding (Section~\ref{sec:termination criteria}). Although some sources exhibit FWHM values larger than 5 pixels (up to $\sim6$ pixels), this does not contaminate the $\mathrm{BI}$ measurement (see upper panel of Figure~\ref{fig:ADU profile}), so that elevated $\mathrm{BI}$ values are driven primarily by the bleeding artifact rather than by the star itself. 

\begin{figure*}[t]
\centering
\includegraphics[width=0.8\textwidth]{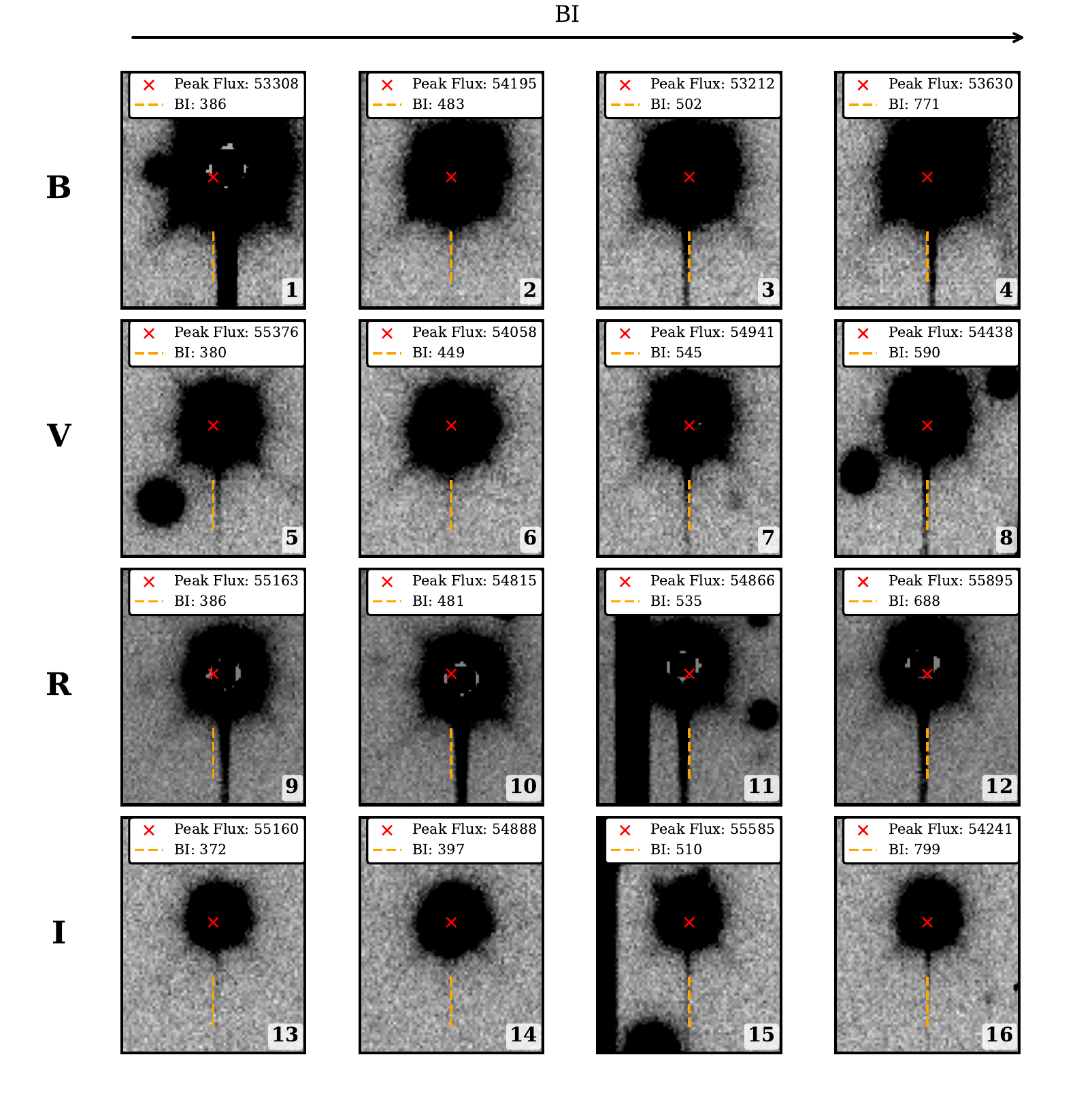}
\caption{$BVRI$ images of sources with increasing $\mathrm{BI}$ values from left to right. In each panel, the candidate bleed-generating pixel (peak flux) is marked with a red cross, and the 20-pixel segment used to compute $\mathrm{BI}$ is indicated by a yellow dashed line. The 20-pixel segments for sources with $\mathrm{BI} > 500$ align with the bleed trails.}
\label{fig:BI}
\end{figure*}

We visually inspected bright sources with different $\mathrm{BI}$ values to determine an appropriate threshold for identifying bleed-generating pixels. We used $BVRI$-band images from the N-chip, taken as a representative chip. Figure~\ref{fig:BI} shows four sources per band, with increasing $\mathrm{BI}$ values from left to right. In each panel, the candidate bleed-generating pixel is marked with a red cross, and the 20-pixel segment used to compute $\mathrm{BI}$ is indicated by a yellow dashed line. Sources with $\mathrm{BI} \le 500$ show no clear bleed trails, whereas those with $\mathrm{BI} > 500$ exhibit prominent bleeding features. We therefore adopted $\mathrm{BI}=500$ as the threshold for identifying true bleed-generating pixels. Some sources (labels 1, 5, 9, and 10) have $\mathrm{BI} \le 500$ despite lying close to visible bleed trails. In these cases, the candidate pixels are adjacent to the trail rather than located on the pixels that actually generate the bleeding. This occurs because very bright sources often contain multiple pixels above the bleeding threshold, although only some of them produce bleed trails.

In summary, the algorithm begins masking pixels when the pixel value exceeds $S_{\rm{bl}}$ and $\mathrm{BI} > 500$.

\subsubsection{Detection Threshold \& Continuity Length}\label{sec:termination criteria}

Once the onset of bleeding is identified, its termination must also be determined, since bleeding does not extend along the entire column. We therefore defined a criterion to determine whether each pixel is affected by bleeding. Because bleeding-affected pixels exhibit higher pixel values relative to the local background, the algorithm traces the bleed trail along the column by flagging pixels whose values exceed a detection threshold above the background:
\begin{equation}\label{eq:detection threshold}
    I(x, y) > B + n\sigma_{\rm B},
\end{equation}
where $I(x, y)$ is the pixel value at position $(x, y)$, $B$ is the background level, $\sigma_{\rm B}$ is the background noise, and $n$ is a tunable parameter that controls the mask sensitivity. We estimated $B$ and $\sigma_{\rm B}$ from sigma-clipped statistics of the full-chip image using \texttt{astropy.sigma\_clipped\_stats} with the default settings (clipping threshold = 3.0; maximum iterations = 5), adopting the returned median and standard deviation, respectively.

As a bleed trail extends further, the excess signal diminishes and eventually becomes indistinguishable from the background. To capture this transition, we defined a continuity length, $CL$: masking is terminated when $CL$ consecutive pixels do not satisfy Equation~\ref{eq:detection threshold}. Larger values of $n$ and smaller values of $CL$ yield more conservative (shorter) masks, while smaller $n$ and larger $CL$ extend the masked region at the risk of flagging unaffected pixels.

To determine the optimal combination of $n$ and $CL$, we evaluated each parameter set by comparing the photometry of bleeding-flagged sources against the Gaia Data Release 3 (Gaia DR3; \citealt{gaia2016, gaia2023}) catalog. The key idea is that bleed trails artificially brighten affected sources, producing a systematic negative magnitude offset relative to Gaia. A good mask should flag as many of these photometrically biased sources as possible while minimizing the number of unaffected sources that are incorrectly flagged.

\begin{figure*}[t]
\centering
\includegraphics[width=\linewidth]{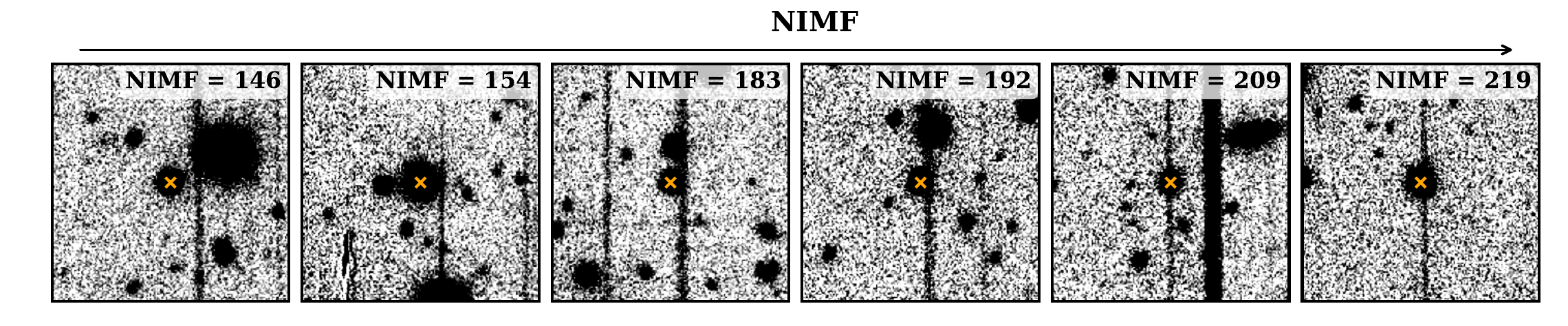}
\caption{Sources are shown as \texttt{NIMAFLAGS\_ISO} increases from left to right. The \texttt{NIMAFLAGS\_ISO} value (\textbf{NIMF}) is indicated in the upper-right corner of each panel, and the target source is marked with a yellow cross. From visual inspection, we confirmed that sources with $\texttt{NIMAFLAGS\_ISO} > 190$ typically lie on bleed trails; we therefore compared their magnitudes with the Gaia DR3 catalog.}
\label{fig:nimf}
\end{figure*}

Candidate masks were generated for a range of $n$ and $CL$ values. First, for each mask, we ran \texttt{SExtractor} with the mask supplied as \texttt{FLAG\_IMAGE} and used \texttt{NIMAFLAGS\_ISO} to quantify the overlap between each detected source and masked pixels. Figure~\ref{fig:nimf} shows six representative sources with increasing \texttt{NIMAFLAGS\_ISO} values from left to right. \texttt{NIMAFLAGS\_ISO} is denoted by $\mathrm{NIMF}$. The target source in each panel is marked with a yellow cross. Although Figure~\ref{fig:nimf} presents only representative examples, we visually inspected a substantially larger sample of sources spanning a broad range of \texttt{NIMAFLAGS\_ISO} values and concluded that sources with $\texttt{NIMAFLAGS\_ISO} > 190$ generally overlap bleed trails sufficiently to be classified as bleed-affected. We therefore adopted $\texttt{NIMAFLAGS\_ISO} > 190$ as the criterion for selecting bleed-affected sources in the following analysis.

Next, we cross-matched these sources with their Gaia DR3 counterparts using a matching radius of $1^{\prime\prime}$ and restricted the sample to high-confidence stellar detections with $\texttt{CLASS\_STAR}>0.9$ and $\texttt{FLAGS}<4$, for which Gaia photometry is expected to be reliable. The $\texttt{FLAGS}<4$ criterion excludes severely blended or otherwise problematic detections while retaining mildly affected sources (typically with $\texttt{FLAGS}=1$--2), which remain relevant for evaluating mask performance. For each matched source, we compared a synthetic $I$-band magnitude from the Gaia XP spectrum \citep{gaia2023} with the KMTNet $I$-band \texttt{MAG\_AUTO} measurement.

We then classified each source according to the difference between the KMTNet and Gaia-based magnitudes. A source was classified as \textit{Goodphot} if the magnitude difference was consistent with zero within the combined uncertainty. A source was classified as \textit{Badphot} if the KMTNet $I$-band magnitude was brighter than the Gaia synthetic magnitude by more than the combined uncertainty. The combined uncertainty was defined as
\begin{equation}
    \sigma_{\mathrm{combined}} = \sqrt{\texttt{MAGERR\_AUTO\_I}^{2} + \sigma_{\mathrm{Gaia\ XP}}^{2}}.
\end{equation}

Finally, to identify the best-performing mask, we sought a configuration that maximizes the number of flagged Badphot sources while minimizing the number of flagged Goodphot sources. To this end, we defined a score that combines two quantities: the purity of the flagged sample and the relative recovery of Badphot sources. The score has an $F_{\beta}$-like form,
\begin{equation}
    \mathrm{Score} \equiv \frac{(1+\beta^{2})\,P\,R_{\rm rel}}{\beta^{2}P + R_{\rm rel}},
\end{equation}

where
\begin{equation}
    P = \frac{N_{\mathrm{Badphot}}}{N_{\mathrm{Goodphot}} + N_{\mathrm{Badphot}}},
\end{equation}

and
\begin{equation}
    R_{\rm rel} = \frac{N_{\mathrm{Badphot}}} {\max_{j}\!\left(N_{\mathrm{Badphot},j}\right)}.
\end{equation}
Here, $P$ is the purity of the flagged sample, i.e., the fraction of flagged
sources that are classified as Badphot, and $R_{\rm rel}$ is a relative recovery
metric defined as the number of recovered Badphot sources normalized by the
maximum value obtained among the candidate masks tested. Because the total
number of truly bleed-contaminated sources is not known a priori, an absolute
recall cannot be computed. Accordingly, $R_{\rm rel}$ should be interpreted only
as a relative measure for comparing candidate masks, not as an absolute
recovery fraction. The parameter $\beta$ controls the relative emphasis on $P$ and $R_{\rm rel}$: $\beta \gg 1$ favors $R_{\rm rel}$, whereas $\beta \ll 1$ favors $P$. We chose $\beta = 0.5$ (i.e., $\beta^{2}=0.25$) to place greater weight on purity, because the cost of falsely masking an unaffected source is higher than that of missing a \texttt{Badphot} source. Once an unaffected source is masked, it is effectively removed from subsequent analyses. By contrast, a \texttt{Badphot} source that is not flagged by the mask can still be identified and excluded later through quality-control procedures such as photometric outlier rejection.

Using a stacked $I$-band image of the 1084 field, we generated candidate masks for selected combinations of detection threshold ($n=0.3$--1.4) and continuity length ($CL=5$, 6, and 8). Masks were produced for each of the four chips individually and then stacked. Generating a mask for a single chip typically took $\sim$1\,min on a Linux server (Intel Xeon E5-2650 v4, single CPU core). Table~\ref{tab:bmask_resulttab} lists the score for each parameter set. We adopted $n=0.4$ and a continuity length of 6, which yielded the highest $F_{\beta}$-like score. 

\begin{table}
\centering
\captionsetup{justification=raggedright, singlelinecheck=false}
\caption{Number of Badphot and Goodphot sources, and corresponding scores for different masks. Masks were parameterized by the detection threshold (\textit{n}$\sigma$) and the continuity length (\textit{CL}). Rows are grouped by continuity length (5, 6, 8). The selected mask is highlighted in gray.}
\label{tab:bmask_resulttab}
\begin{tabular}{ccccc}
\hline
\textbf{\textit{n}} & \textbf{\textit{CL}} & \textbf{Badphot} & \textbf{Goodphot} & \textbf{Score} \\
\hline
0.5  & 5 & 167 &  77 & 0.640 \\
0.8  & 5 & 153 &  57 & 0.654 \\
1.0 & 5 & 120 &  46 & 0.604 \\
1.2 & 5 & 125 &  42 & 0.627 \\
1.4 & 5 & 114 &  37 & 0.611 \\
\hline
0.3  & 6 & 241 & 143 & 0.646 \\
\rowcolor{gray!30}
0.4  & 6 & 242 & 119 & 0.682 \\
\hline
0.3  & 8 & 328 & 226 & 0.644 \\
0.4  & 8 & 296 & 190 & 0.651 \\
0.5  & 8 & 266 & 160 & 0.654 \\
0.8  & 8 & 212 & 106 & 0.662 \\
1.0 & 8 & 175 &  78 & 0.652 \\
1.2 & 8 & 160 &  64 & 0.653 \\
1.4 & 8 & 141 &  51 & 0.643 \\
\hline
\end{tabular}
\end{table}

\begin{figure*}[t!]
    \centering
    \includegraphics[width=0.49\linewidth]{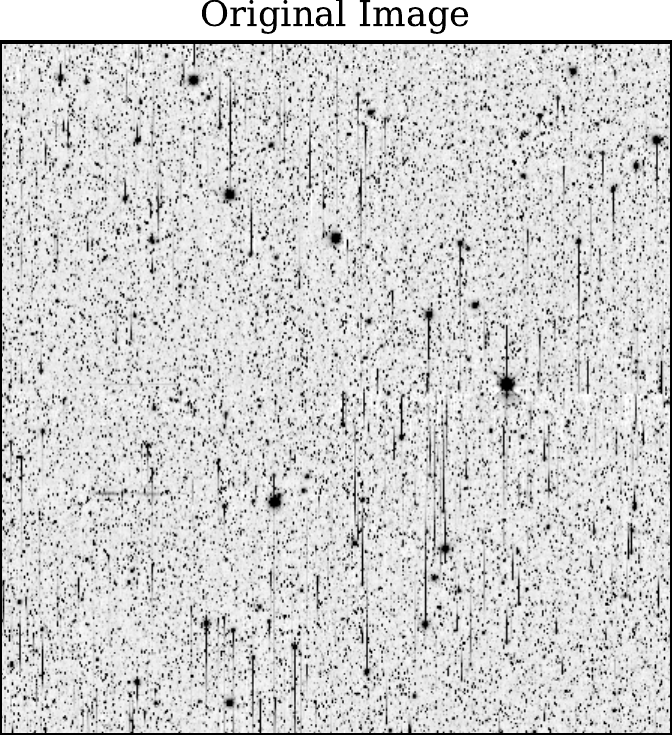}
    \includegraphics[width=0.49\linewidth]{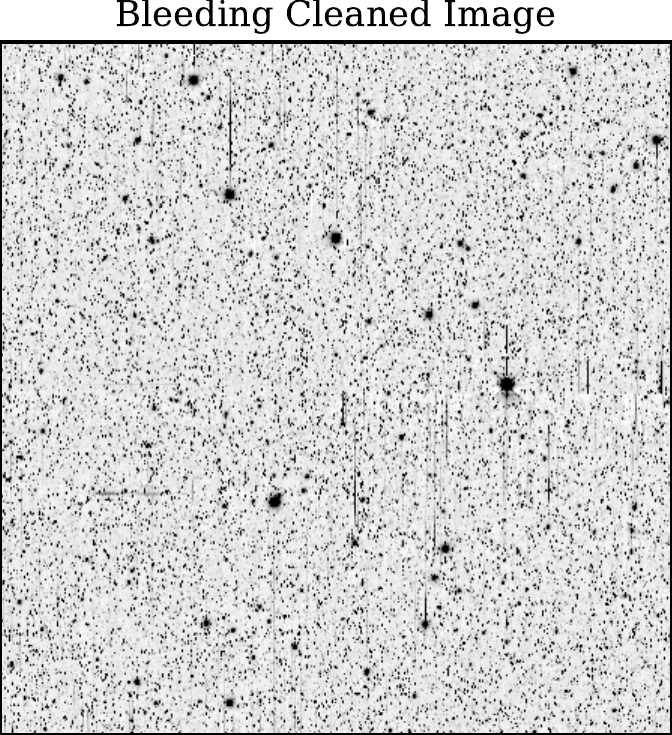}
    \caption{Stacked images combining the K-, M-, T-, and N-chips (left) and the corresponding bleed-cleaned image after interpolation (right). The remaining column-aligned artifacts in the right panel are residual bleed trails after interpolation (Section~\ref{sec:residual bleeding}) or crosstalk \citep{crosstalk}.}
    \label{fig:bleeding comparison}
\end{figure*}

\section{Application}\label{sec:Results}

We applied the algorithm to the 1084 field to generate bleeding masks for each single-epoch image. The mask was constructed using a bleeding threshold of 50,000 ADU (40,000 ADU for the T chip at CTIO), $\mathrm{BI}=500$, $n\sigma=0.4\sigma$, and $CL=6$. We used these masks to correct bleeding with \texttt{SExtractor}, following the procedure in \citet{jeong2026}. First, individual masks were combined into a master bleeding mask covering the full field. The master mask was then inverted so that valid pixels and masked pixels were assigned values of 1 and 0, respectively. This was used as the \texttt{WEIGHT\_MAP} in \texttt{SExtractor} for the stacked image of the 1084 field. We set \texttt{MASK\_TYPE}=\texttt{CORRECT} and \texttt{CLEAN}=\texttt{Y} in \texttt{SExtractor} so that pixels flagged by the bleeding mask were replaced by interpolated values estimated from the surrounding valid pixels, while spurious detections were suppressed. To generate the corrected image, we used \texttt{CHECKIMAGE\_TYPE}=\texttt{-BACKGROUND} together with \texttt{BACK\_TYPE}=\texttt{MANUAL} and \texttt{BACK\_VALUE}=0. With this configuration, the \texttt{CHECKIMAGE} \texttt{-BACKGROUND} is identical to the input science image except in the masked regions, where the original bleeding-affected pixels are replaced by interpolated background-like values. We therefore adopted this \texttt{-BACKGROUND} check image as the final bleed-corrected science image. Figure~\ref{fig:bleeding comparison} compares the original stacked image (left) with the bleed-corrected image (right).

With the corrected image, we assessed the improvement in source recovery and photometric accuracy, as these are the two principal issues caused by bleeding. This was done by comparing the catalogs extracted from the original and corrected images of the 1084 field against Gaia DR3.

\begin{figure*}[t!]
  \centering
  \includegraphics[width=0.9\textwidth]{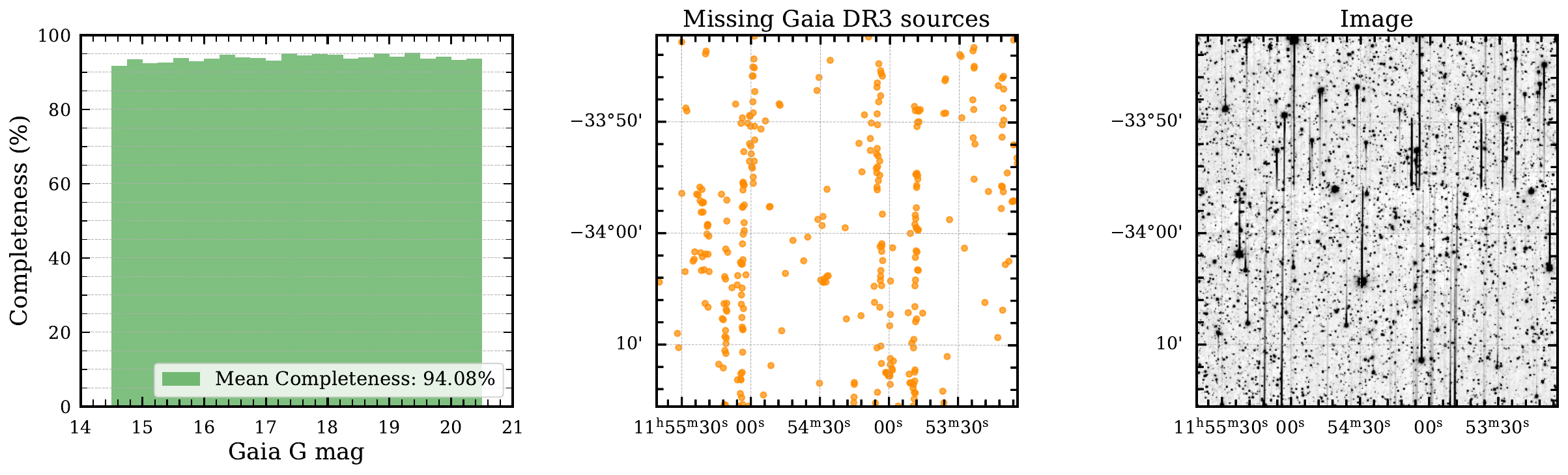}\par
  \vspace{0cm}
  \includegraphics[width=0.9\textwidth]{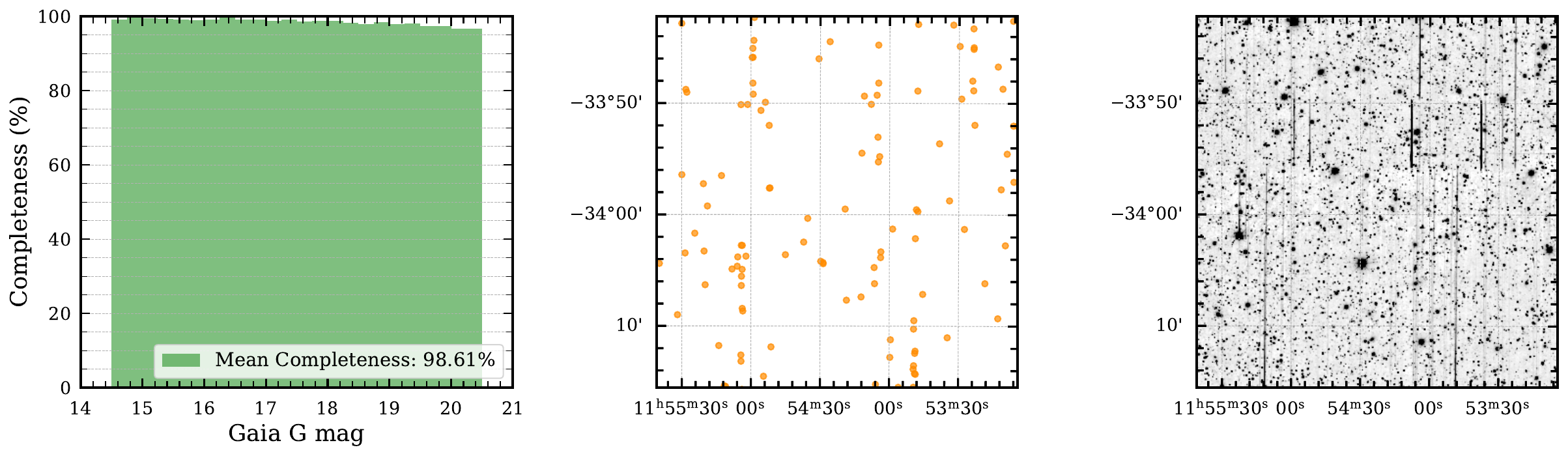}\par
  \caption{Source completeness of the 1084 field in the $I$-band, evaluated relative to Gaia DR3. The upper row shows results from the original image, and the lower row shows results from the bleed-cleaned image. The left panels present completeness as a function of Gaia $G$ magnitude over $14.5 \le \mathrm{Gaia}\ G \le 20.5$. The middle panels mark Gaia sources without a matched KMTNet counterpart in a representative subregion, and the right panels show the corresponding image cutouts.}
  \label{fig:completeness}
\end{figure*}

\subsection{Completeness}\label{sec:completeness}

We evaluated the source completeness of the 1084 field by measuring the fraction of Gaia DR3 sources recovered in the KMTNet catalogs as a function of magnitude. Over the field and magnitude range considered here, Gaia DR3 is effectively complete for our purpose, and thus provides a robust reference set of sources. The Gaia completeness in this field was estimated using \texttt{m10\_to\_completeness} in the \texttt{gaiaunlimited.selectionfunctions} package. The function \texttt{m10\_to\_completeness} converts the local $M_{10}$ value into the predicted Gaia DR3 completeness at a given $G$ magnitude. Here, $M_{10}$ is defined as the median $G$ magnitude of sources with \texttt{astrometric\_matched\_transits} $\leq 10$ in a given sky region, and serves as a proxy for the local survey depth. Following the empirical Gaia DR3 selection-function model of \citet{cantatgaudin2023}, the completeness is thus estimated as a function of both magnitude and sky position through the local $M_{10}$ value. Before cross-matching, we corrected the Gaia positions for proper motion. The Gaia DR3 positions are given at the reference epoch J2016.0, whereas the KMTNet source positions were measured from images obtained in 2021 and expressed in the J2000.0 coordinate system. We propagated each Gaia DR3 source to the epoch of the KMTNet image using its proper-motion measurements, and then cross-matched the original-image and bleed-cleaned catalogs to the Gaia DR3 catalog using a matching radius of $1.5^{\prime\prime}$. This correction is necessary to avoid mismatches for high-proper-motion stars: over the approximately five-year epoch difference, sources with proper motions larger than about 300 mas yr$^{-1}$ would shift by more than the adopted matching radius. In our sample, three stars have proper motions above this threshold, illustrating that the correction can affect individual cross-matches even when the number of such sources is small. The analysis was restricted to $14.5 \le \mathrm{Gaia}\ G \le 20.5$, excluding bright saturated stars and limiting the sample to the magnitude range in which Gaia is effectively complete.

\begin{figure}[t!]
    \centering
    \includegraphics[width=0.7\linewidth]{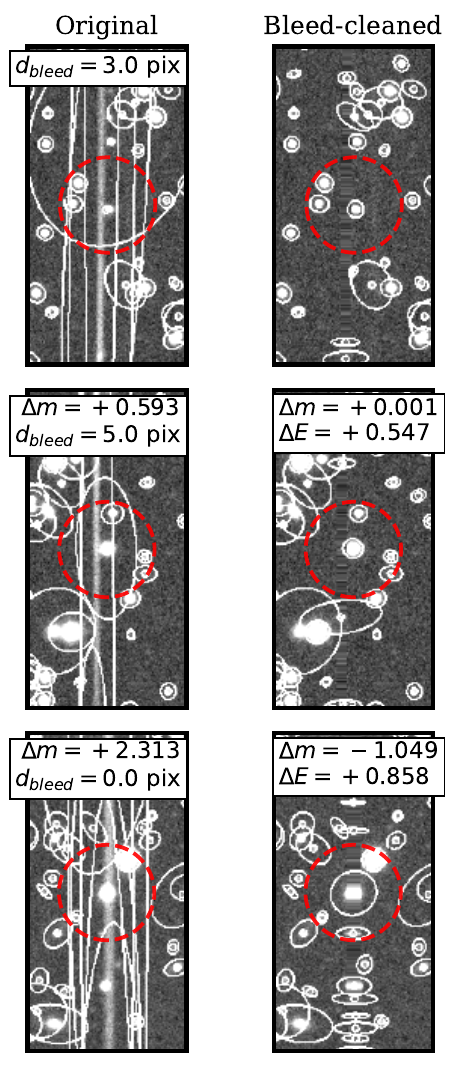}
    \caption{\texttt{APERTURES} \texttt{CHECKIMAGE}s from \texttt{SExtractor} for three representative sources. In each row, the left and right panels correspond to the original and bleed-cleaned images, respectively, and the red dashed circle marks the target source. The top row shows a source recovered near a bleed trail, demonstrating improved completeness through better aperture placement. The middle row shows a source whose photometry improves after bleed cleaning. The bottom row shows a source located directly on a bleed trail; in this case, the photometry becomes worse after cleaning because interpolation replaces bleed-dominated pixels within the source footprint. The distance along \texttt{X\_IMAGE} from each object to the bleed trail is also shown.}
    \label{fig:apertures}
\end{figure}

Figure~\ref{fig:completeness} shows the completeness of the original image (upper) and the bleed-cleaned image (lower). The left panels show the completeness as a function of Gaia $G$ magnitude. Completeness increased in all adopted magnitude bins after bleeding correction, with the mean completeness rising from 94.08\% to 98.61\%. The middle panels present Gaia sources lacking a matched KMTNet counterpart in a representative subregion, while the right panels provide the corresponding image cutouts. Before bleeding correction, many of the unmatched sources are spatially coincident with bleed trails, indicating that bleeding substantially degrades source detection. After cleaning, this concentration is greatly reduced. An example of a source recovered after bleeding correction is shown in the top row of Figure~\ref{fig:apertures}. No aperture was placed in the original image (left), whereas the source is properly detected and assigned an aperture in the bleed-cleaned image (right).

Some sources remain unmatched even after cleaning. Most of these lie directly on severe bleed trails, where the interpolation suppresses a substantial fraction of the source flux, making the source undetectable. Others are associated with residual bleeding left by the current interpolation scheme; this limitation is discussed in Section~\ref{sec:residual bleeding}. A smaller fraction of unmatched sources is attributable to other image artifacts, such as diffraction spikes, or to deblending failures due to nearby sources.

\subsection{Photometry}\label{sec:photometry}

Bleeding correction can improve photometry; however, this does not apply to all affected sources. Improvement is expected primarily for sources located close enough to a bleed trail for their segmentation or aperture to be distorted, but not so close that a substantial fraction of their intrinsic flux is replaced during interpolation. In contrast, sources that overlap the bleed trail itself can become fainter after cleaning, because bleed-dominated pixels within the source footprint are replaced with background-like values.

To isolate the regime in which photometry is expected to improve, we ran \texttt{SExtractor} on both the original and the bleed-cleaned images using the bleeding mask as the \texttt{FLAG\_IMAGE}. Output catalogs were then cross-matched to each other. We examined the change in source shape between the two KMTNet catalogs to identify objects whose photometric apertures were likely distorted by nearby bleed trails. Specifically, we adopted
\begin{equation}
    \Delta \texttt{E}
    \equiv
    \texttt{ELLIPTICITY}_{\rm orig}
    -
    \texttt{ELLIPTICITY}_{\rm clean},
\end{equation}
\begin{equation}
    \Delta \texttt{ELLIPTICITY} > 0.1.
\end{equation}
Here, \texttt{ELLIPTICITY} is the \texttt{SExtractor} shape parameter derived from the source semi-major and semi-minor axes. This threshold was chosen empirically through visual inspection and preferentially selects sources whose segmentation is elongated along the bleed trail in the original image. We further required \texttt{NIMAFLAGS\_ISO} $> 0$, ensuring that at least one pixel in the source isophotal footprint overlaps the bleeding mask, and is therefore expected to show improved photometry.

\begin{figure}[t!]
    \centering
    \includegraphics[width=\linewidth]{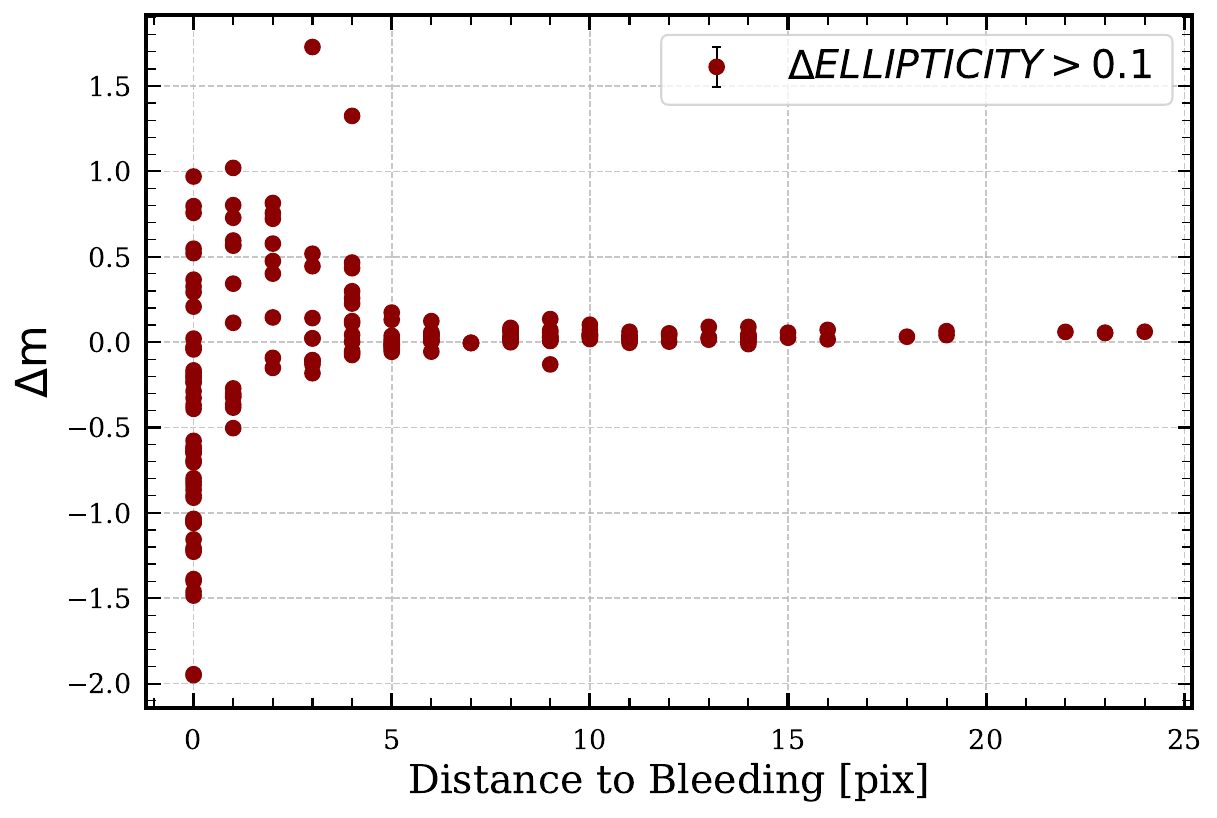}
    \caption{$\Delta m$ of sources in the bleed-cleaned image, as a function of distance from the bleeding mask. The samples are restricted to sources with $\Delta\texttt{ELLIPTICITY} > 0.1$ between the original and bleed-cleaned images and $\texttt{NIMAFLAGS\_ISO} > 0$.}
    \label{fig:magdiff_disttobleed}
\end{figure}

We then compared the cross-matched catalog with the synthetic $I$-band magnitudes derived from the Gaia XP spectra to assess whether the photometry of bleeding-affected sources improved. Before doing so, we excluded sources whose own flux is strongly altered by the interpolation, because improvement is not expected in such cases. We therefore required that no bleeding-masked pixels be present within $\pm 5$ pixels in the \texttt{X\_IMAGE} direction. This threshold was motivated by the behavior of the magnitude offset,
\begin{equation}
    \Delta m \equiv m_{{\rm Gaia\,XP},I} - \texttt{MAG\_AUTO\_I},
\end{equation}
in the bleed-cleaned image as a function of distance from the bleeding mask along the $x$-axis. As shown in Figure~\ref{fig:magdiff_disttobleed}, sources at smaller separations exhibit large offsets whereas sources located more than 5 pixels away show offsets within 0.2 mag. This indicates that direct contamination by the bleed trail is significantly reduced if it is located more than five pixels away from the source center. This five-pixel threshold also matches the median FWHM of unsaturated stars as discussed in Section~\ref{sec:BI}. Finally, we restricted the sample to $14 < \texttt{MAG\_AUTO\_I} < 21$ to exclude saturated sources and objects near the KMTNet limiting magnitude \citep{chang2026}. In total, 125 sources satisfied these criteria.

Examples of a source located adjacent to a bleed trail and a source lying directly on a bleed trail are shown in the second and third rows of Figure~\ref{fig:apertures}, along with their magnitude offsets and distances to the bleeding mask. As expected, the source in the middle row appears artificially bright before cleaning, but agrees much better with the Gaia XP synthetic magnitude after cleaning. In contrast, the source in the bottom row becomes fainter after cleaning, reflecting the removal of bleed-dominated pixels from within the source footprint.

\begin{figure}
    \centering
    \includegraphics[width=\linewidth]{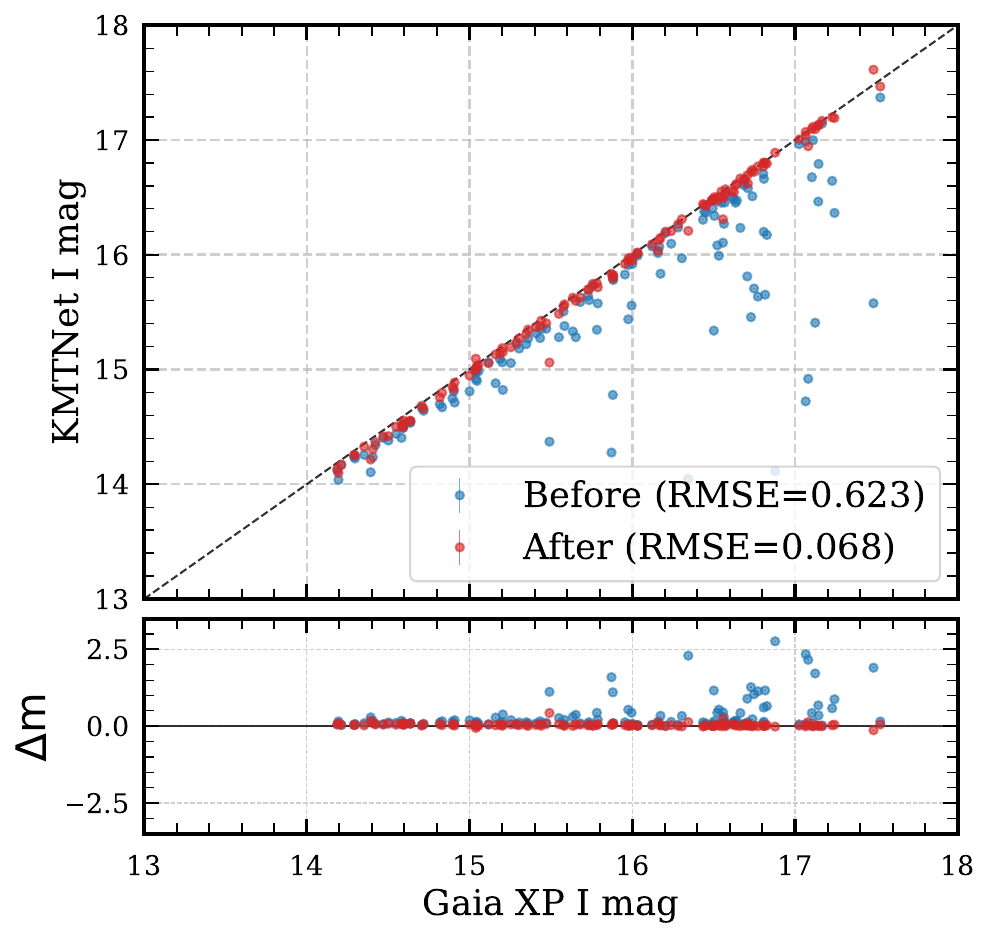}
    \caption{Comparison of KMTNet $I$-band \texttt{MAG\_AUTO} and synthetic $I$-band magnitudes derived from the Gaia XP spectra. The upper panel shows the photometric comparison, and the lower panel shows the offset $\Delta m$. Blue and red points indicate measurements from the original and bleed-cleaned images, respectively. The samples are restricted to sources satisfying $\Delta\texttt{ELLIPTICITY} > 0.1$ and the additional selection criteria described in Section~\ref{sec:photometry}.}
    \label{fig:phot}
\end{figure}

Figure~\ref{fig:phot} summarizes the photometric improvements after bleed-cleaning. For the selected sample, the distribution of $\Delta m$ becomes markedly more concentrated around zero after cleaning. The root-mean-square (RMS) error in $\Delta m$ decreases from 0.623 mag in the original image to 0.068 mag, demonstrating a substantial improvement in the agreement between KMTNet photometry and the Gaia XP-based reference values.

\section{Discussion}\label{sec:Discussion}

Here we evaluate whether the masking algorithm, developed and tuned using a single-epoch CTIO $I$-band image, can be applied reliably to data obtained at other sites, in other bands, and under different observing conditions. We focus mainly on the stability of the $\mathrm{BI}$ criterion. We also discuss the limitations of the current algorithm and the bleed-cleaning process.

\subsection{Stability of the Bleeding Index Criterion}

As defined in Equation~\ref{eq:BI equation}, $\mathrm{BI}$ is computed relative to the local background level and is therefore sensitive to background variations. In addition, poor seeing may broaden the source profile enough for source flux to contribute to $\mathrm{BI}$ at the sampled pixels. Because both the background level and seeing vary with site, filter, and lunar phase, it is necessary to test whether the fixed threshold adopted in this work ($\mathrm{BI}>500$) remains applicable under a range of observing conditions.

To examine the effect of the background level, we analyzed 24 representative images spanning three sites (SAAO, SSO, and CTIO), four bands ($BVRI$), and two lunar conditions (full moon and dark time). For each image, we measured the background level and visually inspected bright sources to assess whether the adopted threshold consistently suppresses false bleeding detections. Across this sample, the background level spans a wide range, from $\sim$60 to $\sim$3600 ADU. Despite this variation, the criterion showed consistent performance in our test images, with no clear evidence that the fixed threshold systematically breaks down at either low or high background levels. This suggests that the adopted $\mathrm{BI}$ threshold is sufficiently stable for practical use across a broad range of sky backgrounds.

\begin{figure*}[t!]
    \centering
    \includegraphics[width=\linewidth]{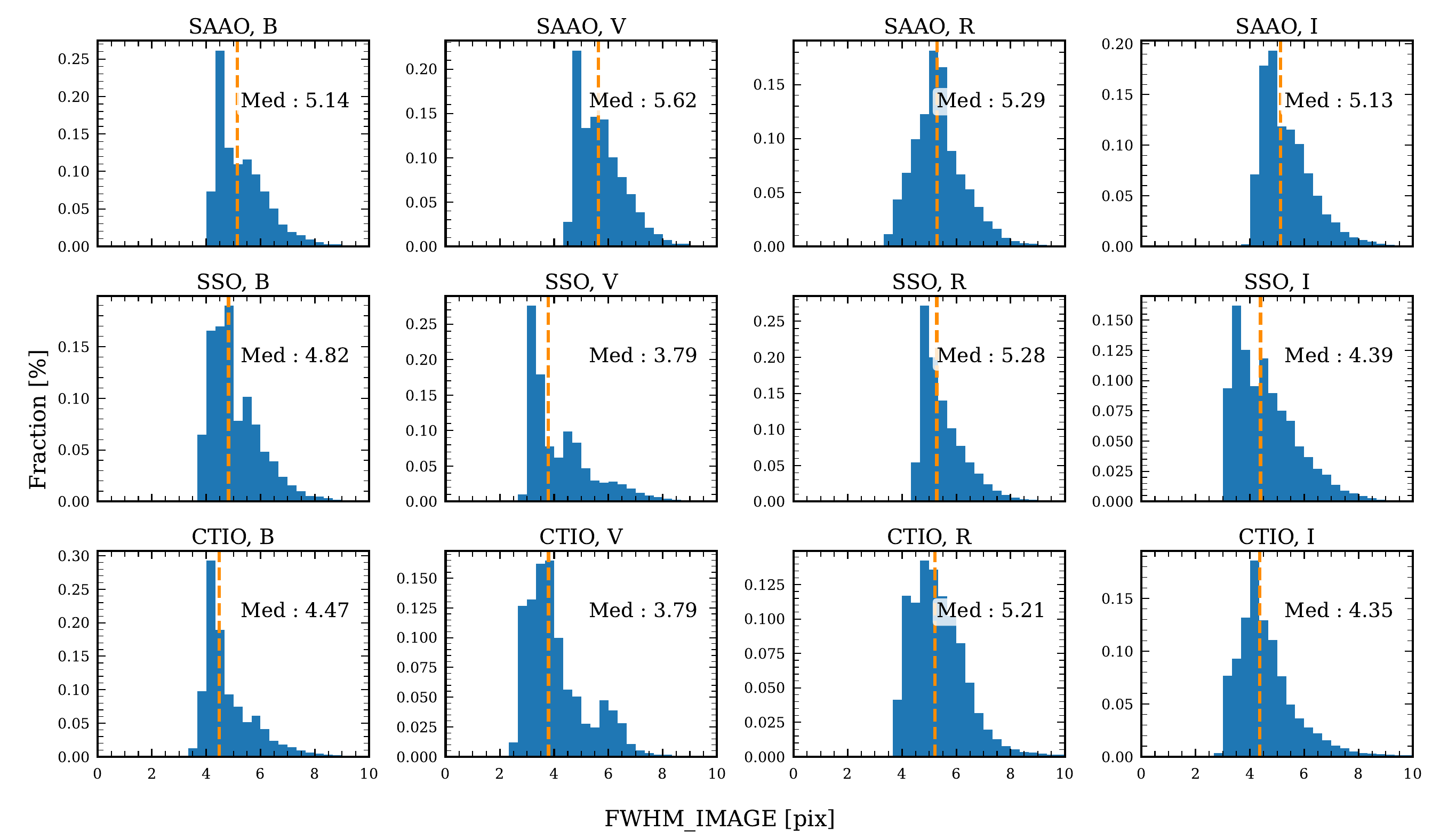}
    \caption{FWHM distributions of unsaturated, bleed-unaffected sources derived from five images for each site (SAAO, SSO, and CTIO) in each of the $B$, $V$, $R$, and $I$ bands. The orange dashed line in each panel indicates the median FWHM.}
    \label{fig:fwhm_hists}
\end{figure*}

We next examined the effect of seeing using 60 images, corresponding to five images from four bands across the three sites. For each image, we measured the FWHM of unsaturated point sources selected using the same criteria as in Section~\ref{sec:BI} ($\texttt{CLASS\_STAR}>0.9$, $\texttt{NIMAFLAGS\_ISO}<190$, and $\texttt{FLAGS}=0$). Figure~\ref{fig:fwhm_hists} presents the FWHM distributions and their median values for all site--band combinations. The median FWHM varies by at most 1.8 pixels across the sample, with the largest value in the SAAO $R$-band image and the smallest values in the SSO $V$-band and CTIO $V$-band images. This variation is modest compared to the 20-pixel offset at which $\mathrm{BI}$ is evaluated from the peak pixel. Furthermore, we confirmed that even for sources with unusually large FWHM values, the $\mathrm{BI}$ criterion still holds (see upper panel of Figure~\ref{fig:ADU profile}). Therefore, the contribution of the broadened PSF wings to $\mathrm{BI}$ is expected to remain small, and the adopted starting point for the $\mathrm{BI}$ measurement should remain valid over the range of seeing conditions considered here.

Taken together, these tests indicate that the $\mathrm{BI}$, although calibrated using a single CTIO $I$-band image, is not strongly sensitive to plausible variations in background level or image quality. We therefore conclude that the criterion is reasonably transferable across the KMTNet sites and $BVRI$ bands. However, this conclusion is based on a limited validation set and primarily on visual inspection, so a larger-scale statistical assessment would be desirable in future work.

\subsection{Limitations}\label{limitation}

\subsubsection{Artifacts Introduced by Interpolation}
In the right-hand panel of the bottom row of Figure~\ref{fig:apertures}, numerous sources are detected along the interpolated bleed trail, despite being absent in the original image. These detections are artifacts introduced by the interpolation step. In \texttt{SExtractor}, masked pixels are filled by copying the value of the nearest unmasked pixel to the left (within \texttt{INTERP\_MAXXLAG}), so horizontally adjacent pixel values can be copied over extended regions. When the values lie above the local background, interpolated pixels can be mistakenly detected as real sources. The resulting apertures are typically elongated along the $x$-axis and align with the bleeding mask, reflecting their non-astrophysical origin. This is particularly problematic for time-domain applications when using the cleaned image, where spurious detections may contaminate transient searches.

To mitigate this issue, an alternative replacement scheme for bleeding-masked pixels could be applied. Rather than copying the value of an adjacent pixel, masked pixels could be filled with background-like values drawn from a Gaussian distribution with mean equal to the sigma-clipped background level and standard deviation equal to the local background RMS. This approach would suppress coherent structures created by replication, reduce false detections, and produce visually more natural backgrounds. Another simple solution would be to exclude sources whose positions lie within five pixels of the bleeding mask along the \texttt{X\_IMAGE} direction, as their photometry is likely to be unreliable (Figure~\ref{fig:magdiff_disttobleed}).

\begin{figure*}
    \centering
    \includegraphics[width=\linewidth]{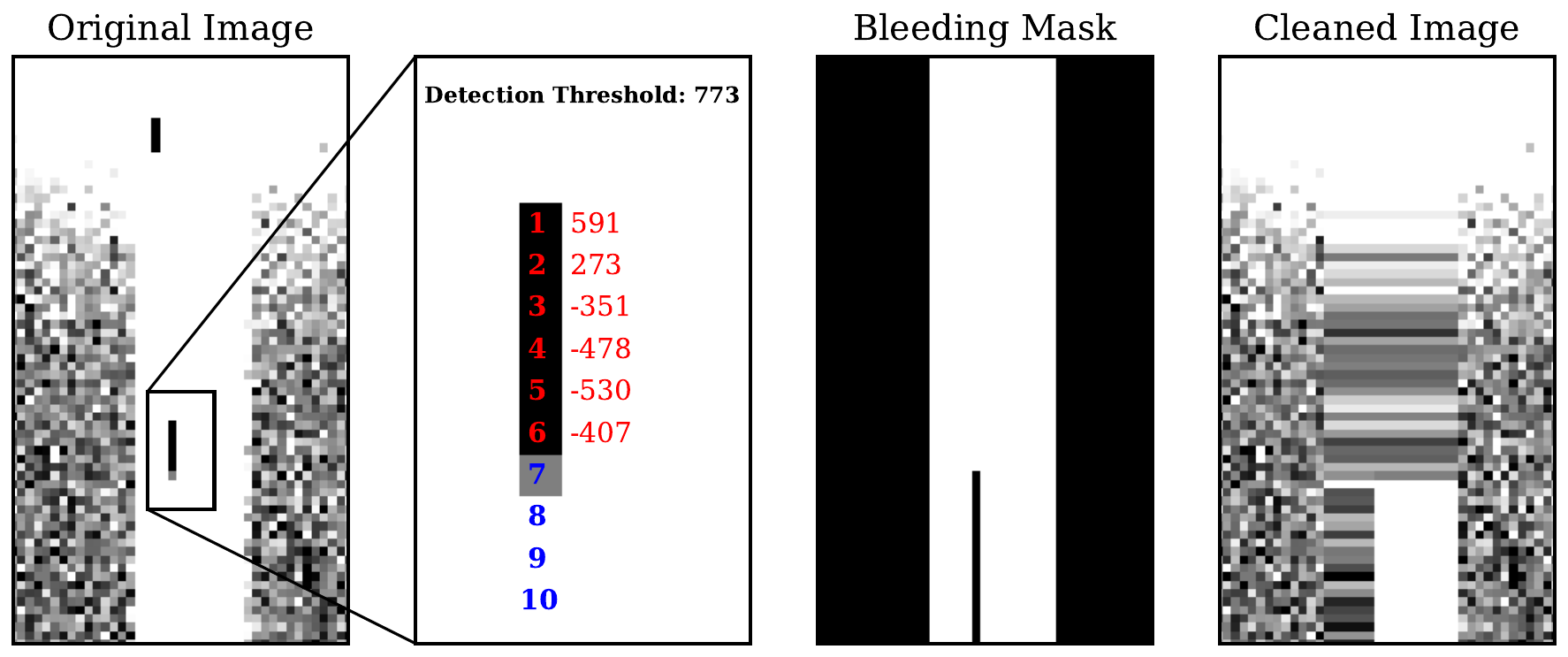}
    \caption{Example of residual bleeding in the cleaned image caused by premature termination of the masking procedure. 
    From left to right, the panels show: (1) a segment of the original bleed trail containing a short sequence of anomalously depressed pixels; 
    (2) the same column with the termination condition satisfied because six consecutive pixels fall below the detection threshold; 
    (3) the resulting incomplete mask, in which subsequent bleeding-affected pixels remain unmasked; and 
    (4) the interpolated image, where those unmasked contaminated pixels bias the interpolation and produce a residual bleeding pattern.}
    \label{fig:limitation}
\end{figure*}

\subsubsection{Residual Bleeding Caused by Incomplete Masking}\label{sec:residual bleeding}

Inspection of the cleaned images showed that some residual bleeding remains even after masking and interpolation (Figure~\ref{fig:bleeding comparison}). Examination of these cases suggests that the residuals are caused by premature termination of the masking procedure. Along the column passing through a bright pixel, we occasionally found short sequences of vertically adjacent pixels with anomalously low values, in some cases even below 0. Although the origin of these depressed pixels is not yet clear, they can satisfy the current termination criterion before the bleed trail has actually ended. As a result, the algorithm stops masking too early, leaving genuinely bleeding-affected pixels below the stopping point unmasked.

Figure~\ref{fig:limitation} illustrates this failure mode. In the first panel, seven consecutive pixels with values significantly lower than those of the surrounding bleeding-affected pixels are visible. The same region is examined more closely in the second panel, where pixels along the column are labeled sequentially from 1 to 10. For pixels 1 to 6, the corresponding pixel values are displayed next to the labels, showing that these six consecutive pixels fall below the detection threshold. Because those pixels do not satisfy Equation~\ref{eq:detection threshold} (\(B+n\sigma_B = 773\) in this example), the termination condition is met at the pixel labeled 6. The algorithm therefore stops at this location, and pixels from label 7 onward are treated as valid data, even though the bleed trail still extends farther along the column. The resulting incomplete masking is shown in the third panel. In the second panel, red labels denote masked pixels, whereas blue labels denote unmasked pixels.

This effect propagates into the interpolation step performed by \texttt{SExtractor}. Pixels beyond label 7 are treated as valid data, even though they remain contaminated by bleeding. Consequently, when masked pixels on the right side of the incompletely masked column are interpolated, pixels beyond label 7 are used as reference values, leaving a bleeding-like structure. The residual pattern seen in the cleaned image is therefore a consequence of incomplete masking, rather than a failure of the interpolation scheme itself. The resulting artifact is shown in the fourth panel of Figure~\ref{fig:limitation}.

One possible way to mitigate this problem is to revise the termination criterion. In the current algorithm, a pixel contributes to termination when its value falls below the upper threshold \(B+n\sigma_B\). Because this is a one-sided condition, a sequence of strongly depressed outliers can be counted as background-like pixels even when the bleed trail is still present. A more robust criterion would instead require the pixel value to be consistent with the background within the threshold, i.e.,
\begin{equation}
    |I-B| < n\sigma_B,
\end{equation}
where \(I\) is the pixel value. Termination would then require \textit{CL} consecutive pixels satisfying this condition. This two-sided criterion is more closely aligned with the intended meaning of termination, namely that the bleeding signal has become indistinguishable from the local background. It should also reduce the chance that anomalously low pixels trigger premature termination, although a full validation of this modification is left for future work.

\section{Conclusion}\label{sec:Conclusion}

Bleeding is a major artifact in KMTNet images that affects source detection, photometry, and transient searches. We developed a pixel-level masking algorithm to identify bleed trails and generate binary masks for affected pixels. The main algorithm parameters were selected using visual inspection and an $F_\beta$-like score designed to favor high-purity recovery of photometrically biased sources. Using Gaia DR3 as an external reference, we demonstrated that bleed masking improves the mean source completeness from 94.08\% to 98.61\% and reduces the RMS photometric offset of selected bleeding-affected sources, measured against Gaia XP synthetic $I$-band magnitudes, from 0.623 mag to 0.068 mag. We also tested the adopted criteria across different chips, bands, sites, and observing conditions, finding that the algorithm remains stable over the range examined. Finally, we identified limitations of the current cleaning procedure, including interpolation-induced artifacts and residual bleeding caused by premature mask termination, and discussed possible improvements such as alternative pixel replacement schemes and a two-sided termination criterion.

An important strength of our method is its simplicity. The algorithm traces bleed trails directly at the pixel level using threshold-based criteria, without requiring prior knowledge of the artifact profile. For example, \citet{waters2020} modeled burn trails by fitting each affected column with an empirical one-dimensional power law. Such approaches can be effective when the underlying artifact profile is well characterized, but they rely on prior knowledge of its functional form. In contrast, our algorithm is model-independent and can therefore be applied even when the detailed behavior of the CCD artifact is not known a priori. As a result, although it was developed for bleeding artifacts in KMTNet images, the same basic approach may also be useful for masking column-aligned artifacts in other instruments and datasets.

We encourage users of KMTNet images to apply this algorithm in their analyses. The bleeding flags have been incorporated into the KS4 DR1 pipeline using the algorithm described here\footnote{\url{https://github.com/jmk5040/KMTNet_ToO.git}}. Documentation for the algorithm, together with an example of its application, is publicly available\footnote{\url{https://github.com/JiseopShin17/Bleeding-Masking-Algorithm}}.


\acknowledgments
This work was supported by the National Research Foundation of Korea (NRF) grants (RS-2025-00573214 \& RS-2026-25490019) funded by the Korean government (MSIT). This research has made use of the KMTNet system operated by the Korea Astronomy and Space Science Institute (KASI) at three host sites: CTIO in Chile, SAAO in South Africa, and SSO in Australia. Data transfer from the host sites to KASI was supported by the Korea Research Environment Open NETwork (KREONET). SWC acknowledges support from the NRF grants funded by the Ministry of Education (RS-2023-00245013) and the MSIT (RS-2026-25489059). 

This work has made use of data from the European Space Agency (ESA) mission
{\it Gaia} (\url{https://www.cosmos.esa.int/gaia}), processed by the {\it Gaia}
Data Processing and Analysis Consortium (DPAC,
\url{https://www.cosmos.esa.int/web/gaia/dpac/consortium}). Funding for the DPAC
has been provided by national institutions, in particular the institutions
participating in the {\it Gaia} Multilateral Agreement.

We thank Dr. Seung-Lee Kim for helpful comments on bleeding, which helped us in the initial stages of this work.

\bibliography{jkas-bleeding-pattern}

\end{document}